\documentclass[aps,twocolumn,superscriptaddress,showpacs]{revtex4-1}
\usepackage{epsfig,color,amssymb,amsbsy,bm}
\usepackage{graphics}

\usepackage{amsmath}  % needed for \tfrac, \bmatrix, etc.
\usepackage{amsfonts} % needed for bold Greek, Fraktur, and blackboard bold
\usepackage{graphicx} % needed for figures
\usepackage{stackengine}
\renewcommand{\bbox}{\boldsymbol }

\newcommand{\isum}{\mathop{\hbox{$\displaystyle\sum\kern-13.2pt\int\kern1.5pt$}}}
\newcommand{\hs}{\hspace*}
\newcommand{\vs}{\vspace*}
\newcommand{\np}{\newpage}

\newcommand{\eref}[1] {(\ref{#1})}
\newcommand{\Eref}[1] {Eq.~(\ref{#1})}
\newcommand{\Fref}[1] {Fig. \ref{#1}}

\newcommand{\be}{\begin{equation}}
\newcommand{\ee}{\end{equation}}
\newcommand{\br}{\begin{eqnarray*}}
\newcommand{\er}{\end{eqnarray*}}
\newcommand{\ba}{\begin{eqnarray}}
\newcommand{\ea}{\end{eqnarray}}
\newcommand{\bp}{\begin{minipage}}
\newcommand{\ep}{\end{minipage}}

\newcommand{\w}{{\omega}}

\newcommand{\Wcm}[2]{
$\rm {#1}\times10^{{#2}}~W/cm^2$}

\begin{document}

\title{Keldysh-Rutherford model for attoclock}

\author{Alexander W. Bray} 
\affiliation{Research School of Physics and Engineering, The
Australian National University, Canberra ACT 0200, Australia\\}
\author{Sebastian Eckart}
\affiliation{Institut f\"ur Kernphysik, Goethe-Universit\"at,
  Max-von-Laue-Str. 1, 60438 Frankfurt, Germany}
\author{Anatoli S. Kheifets}
\affiliation{Research School of Physics and Engineering, The
Australian National University, Canberra ACT 0200, Australia\\}

\date{\today}

\begin{abstract}

We demonstrate a clear similarity between attoclock offset angles and Rutherford 
scattering angles taking the Keldysh tunnelling width as the impact parameter and the
vector potential of the driving pulse as the asymptotic velocity. This simple model 
is tested against the solution of the time-dependent Schr\"odinger equation using
hydrogenic and screened (Yukawa) potentials of equal binding energy. We observe a smooth 
transition from a hydrogenic to `hard-zero' intensity dependence of the offset
angle with variation of the Yukawa screening parameter. Additionally we make comparison 
with the attoclock offset angles for various noble gases obtained with the
classical-trajectory Monte Carlo method. In all cases we find a close correspondence 
between the model predictions and numerical calculations. This suggests a
largely Coulombic origin of the attoclock offset angle and casts further doubt on its 
interpretation in terms of a finite tunnelling time.

\end{abstract}


\maketitle 

%{\small \tableofcontents}\ns

\section{Introduction}

Measuring an offset angle of the peak photoelectron momentum
distribution in the polarization plane of a close-to-circularly polarized
laser field has been used to determine the tunnelling time which
the photoelectron spends under the barrier
\cite{Eckle2008,P.Eckle12052008,Pfeiffer2012,Landsman:14}.  
Once the photoelectron leaves the tunnel at the
peak value of the electric field, the most probable detection
direction will be aligned with the  vector potential at the
instant of tunnelling. This direction is tilted by $90^\circ$ relative
to the electric field  at this instant. A measurable angular
offset from this axis can be converted to the time the
photoelectron spends under the barrier with the conversion rate of 7.4
attoseconds (1~as = $10^{-18}$~s) per $1^\circ$ at 800~nm. Such a
measurement is termed colloquially the attoclock.  A similar reading
can be obtained from an attoclock driven by a very short circularly
polarized laser pulse
\cite{Torlina2015,PhysRevLett.117.023002,PhysRevA.97.013426,0953-4075-50-5-055602,PhysRevA.97.031402}.

These attoclock measurements were made to address the controversy of a
finite tunnelling time which has many decades of history
\cite{RevModPhys.66.217}.  This controversy is yet to be resolved with
many conflicting reports of a finite time
\cite{Landsman20151,PhysRevLett.119.023201} as opposed to zero
tunnelling time
\cite{Torlina2015,PhysRevLett.117.023002,2017arXiv170705445S,PhysRevA.97.031402}.
A promising pathway to resolving this controversy is to examine an
often neglected aspect of the attoclock measurement, the offset angle
induced by the Coulomb field of the ion remainder. In theoretical
calculations based on a numerical solution of the time-dependent
Schr\"odinger equation (TDSE)
\cite{0953-4075-42-16-161001,PhysRevA.85.023428,PhysRevA.89.021402},
the Coulombic and tunnelling components of the offset angle are
inseparable. The Coulombic field's contribution can be switched off
however by replacing the atomic potential with the short-range Yukawa
potential of the same binding strength. This procedure effectively
eliminates the offset angle thus suggesting zero tunnelling time
\cite{Torlina2015,2017arXiv170705445S}. In various classical
(the TIPIS model \cite{PhysRevLett.111.103003},
back-propagation \cite{PhysRevLett.117.023002,PhysRevA.97.013426}, 
CTMC \cite{0953-4075-50-5-055602}) or
semi-classical simulations (the ARM model \cite{Torlina2015}), the
Coulomb field is separable and its effect can be unambiguously
determined.

If the laser field is not very strong and the pulse is
sufficiently short, we can assess the Coulomb component of the offset
angle by considering the motion of the photoelectron in the Coulomb
field alone, decoupling it from the laser field
altogether. Invoking scattering concepts in photoionization studies
has been very fruitful over the years. The earliest and most relevant
example is the Wigner group delay introduced initially for particle
scattering \cite{PhysRev.98.145} but more recently has found wide use in
quantifying photoemission time delay in multi-photon ionization
\cite{RevModPhys.87.765}.  Further examples are
multi-channel scattering theories: the random phase approximation with
exchange \cite{A90} and the convergent close-coupling theory
\cite{Bray2012135}. The former describes electron scattering and
photoionization of the closed-shell atoms while the latter solves the
Coulomb three-body problem and describes single-photon double
ionization of two-electron atomic targets. 
%
%% In both theories, the photoelectron is ejected from the target atom
%% and then interacts with the ion remainder by exchanging energy and
%% angular momentum with it.  At the same time, the total energy of the
%% scattering system and its angular momentum are conserved during the
%% photoionization process.

In the present work, we apply a similar concept: the photoelectron is
tunnel ionized and then scatters elastically on the ion remainder. We
examine cases where the applied field is relatively weak and the pulse is
short such that the actual trajectory of the field driven photoelectron is
similar to that of an elastically scattered particle.  This allows us
to estimate the offset angle from the classical scattering
formula \cite{Landau1982}
\be
\label{elastic}
\theta = 2\int\limits_{r_0}^\infty
{
(\rho/r^2)\;dr 
\over
[1-(\rho/r)^2-(2V/mv_\infty^2)]^{1/2}
}
-\pi \ .
%\vs{-5mm}
\ee
Here $\rho$ is the impact parameter, $v_\infty$ is the velocity of the
projectile at the source and the detector, and the point of the
closest approach $r_0$ is the largest positive root of of the
denominator.  In the following we adopt the system of atomic units and set the
charge and mass of the electron as well as the reduced Planck constant to
unity $e=m=\hbar=1$.  In the case of the attractive Coulomb potential
$V(r)=-Z/r$,  \Eref{elastic} takes the form of the Rutherford
formula \cite{Landau1982}
\be
\tan{\theta\over2} = {1\over v_\infty^2}{Z\over \rho}
%  \ ,  
%\theta \simeq   {2\over m v_\infty^2}{Z\over \rho}
\ .
\label{Rutherford}
\ee
In a more general case of a screened Coulomb potential
$V(r)=-Z/r\exp(-r/\lambda)$, the offset angle is given by a modified expression
\cite{PhysRev.99.1287} which is, up to a typically small numerical
correction (see Eq.\ 13 of \cite{PhysRev.99.1287}),
\be
\tan{\theta\over2} = {1\over v_\infty^2}{Z\over \rho}
\exp(-1/z_0)
\ .
\label{Yukawa}
\ee
Here $z_0$ is the root of 
$
y(z) = 1-(\rho/\lambda)^2z^2-(d/\lambda)z\exp(-1/z)
$
and $d=2Z/v_\infty$ is the so-called collision diameter.
If the last term in the right hand side of $y(z)$ can be neglected, 
\Eref{Yukawa} acquires a simple screening exponent $\exp(-\rho/\lambda)$.

\begin{figure}[h]
\vs{-0.4cm}
\hs{-8mm}
\includegraphics[scale=0.21]{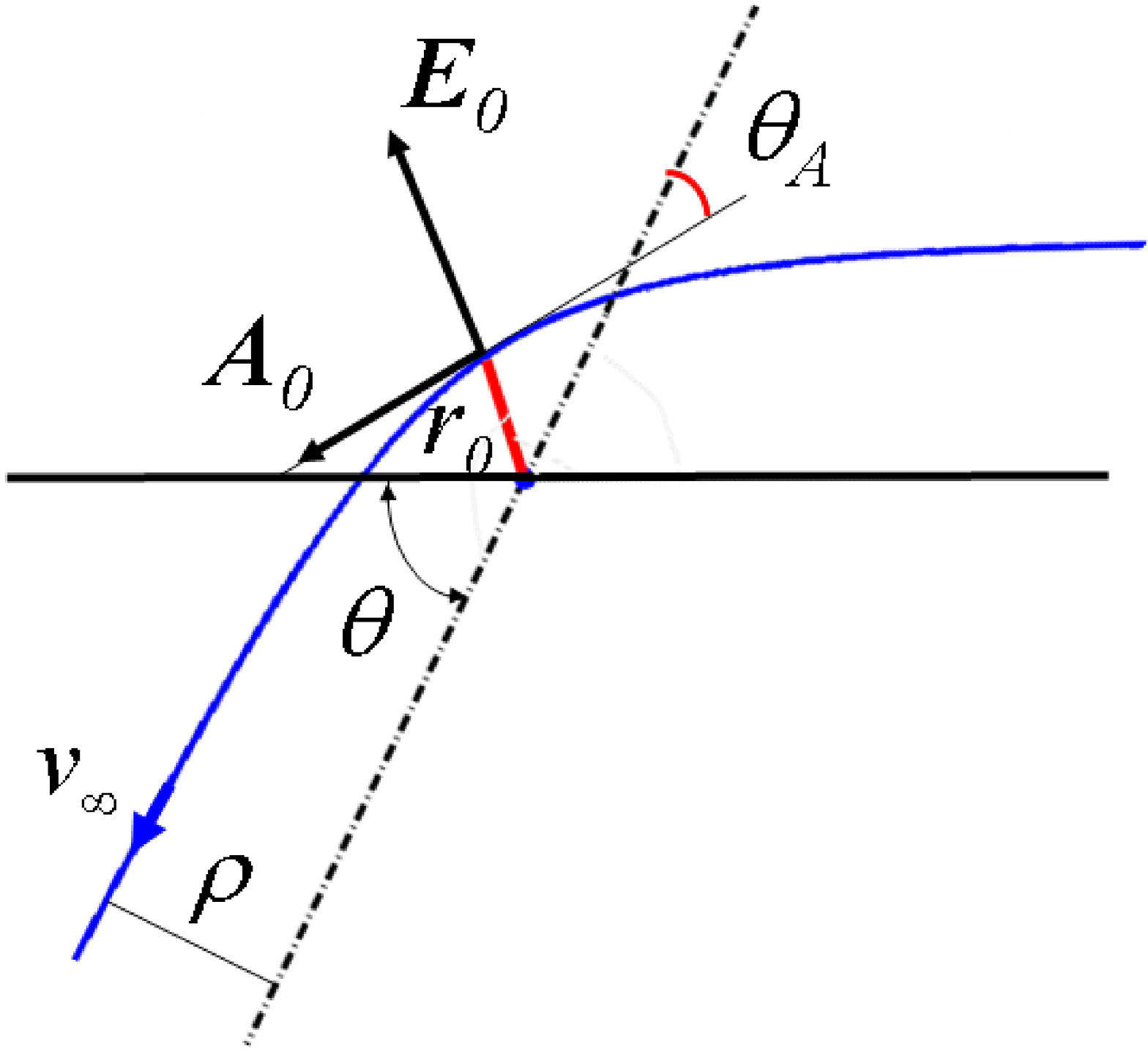}
\vs{2mm}
\includegraphics[scale=0.19]{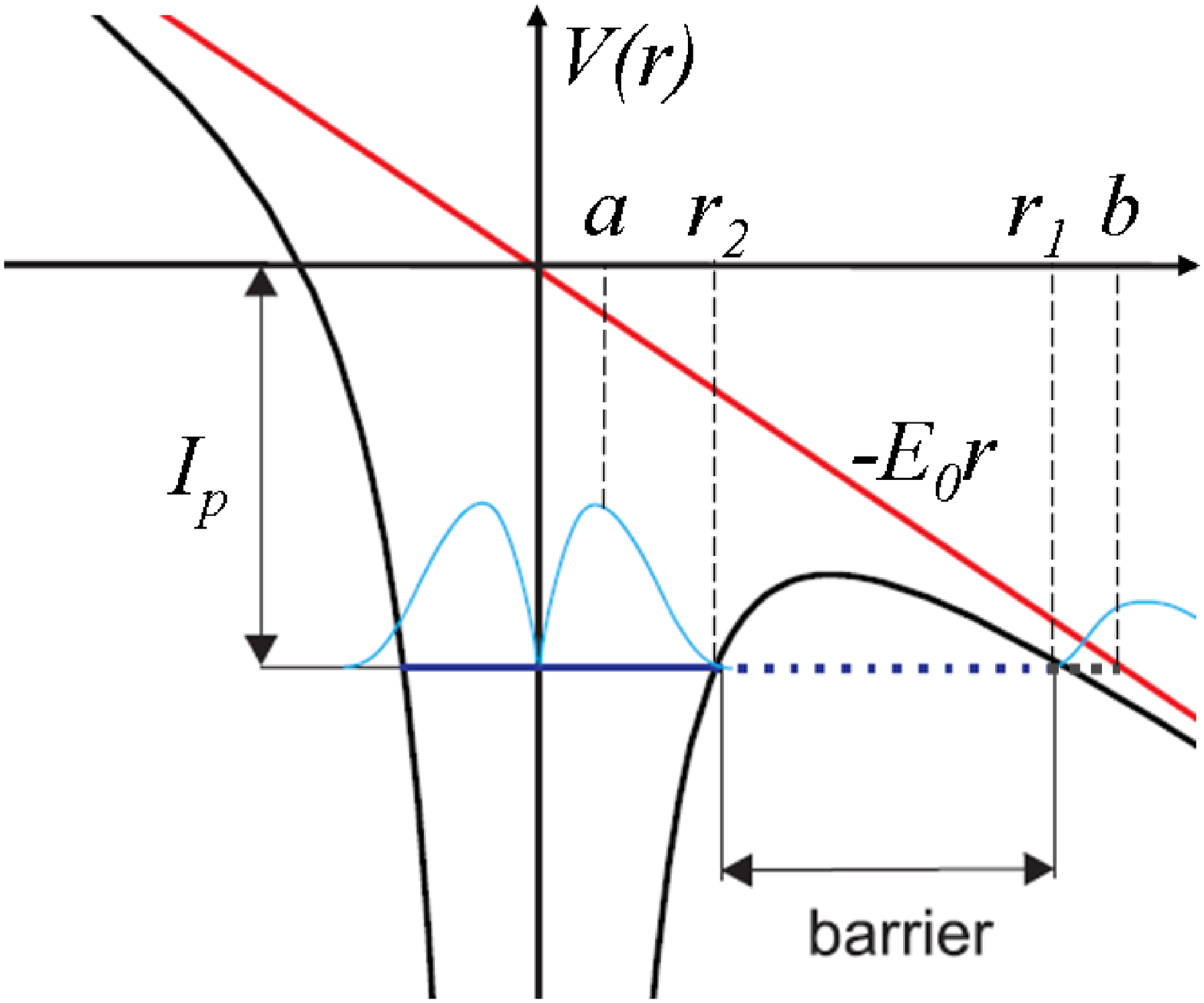}
\vs{-0.4cm}
\caption{ Left: classical scattering trajectory of a particle in a
  central attractive potential. The scattering angle $\theta$ is
  defined by the impact parameter $\rho$ and the asymptotic velocity
  $v_\infty$. The tunnel ionized electron enters this trajectory at
  the point of the closest approach $r_0$ driven by the peak electric
  field ${\bm E}_0$ and arriving to the detector at the angle
  $\theta_A$ relative to the vector potential ${\bm A}_0$.
Right: The Coulomb potential is tipped by the light field. A finite
width potential barrier is created, through which the electron wave
packet leaks out. $I_p$ refers to the binding energy of the electron
in an unperturbed atomic system. See text for further symbol definitions. }
\label{Fig1}
\end{figure}

Eqs.~\eref{Rutherford} and \eref{Yukawa} can be readily applied to the
case of tunnelling  ionization. Firstly, we note that photoionization
can be considered as half-scattering and the offset angle should be
taken as one half of the Rutherford or Yukawa scattering
angles. Similarly, the Wigner time delay in photoemission is equated
with the energy derivative of the scattering phase whereas it is twice
this value in electron scattering. Secondly, the point of the closest
approach $r_0$ should be taken as the tunnel exit position which, for
the Coulomb potential, is the largest root of the equation
\be
Z/r+E_0r = I_p
 \ \ , \ \ 
r_{1,2} = {b/2}\pm\sqrt{b^2/4-ab}
\ .
\ee
Here $I_p$ is the ionization potential, $E_0$ is the peak value of the
electric field, $b=I_p/E_0$ and $a=Z/I_p$. In the weak field limit,
$r_1=b\gg r_2=a\simeq 1$, where $a$ is the characteristic span of the
atomic orbital.  The onset of the over-the-barrier ionization (OBI)
corresponds to $r_1=r_2$ and $E_0= I_p^2/(4Z)$
(right panel of \Fref{Fig1}).
We note that the $b$ parameter is used to evaluate the Keldysh
tunnelling  time
$ 
\tau = b/v_{\rm at} 
$
with
$
v_{\rm at} = \sqrt{2I_p}$ 
which, in turn, defines  the adiabaticity parameter
$
\gamma = \w \tau 
$
\cite{0953-4075-47-20-204001}.
Finally, we equate the asymptotic velocity with the value of the
vector potential at the moment of ionization $v_\infty=A_0$. In the case of adiabatic
tunnelling, the longitudinal velocity of the tunnelled electron in the
direction of the electric field ${\bbox E}_0$ at the point of exit is
zero. Similarly, the radial velocity of the scattering electron at the
point of the closest approach $r_0$ is also vanishing.

To proceed further, we assume that $\rho\simeq r_0$ which is a reasonable
approximation in the case of small scattering angles. This also entails
$\tan(\theta/2)\simeq \theta/2$. 
In addition to this we neglect the influence of the central potential at 
the tunnel exit and simply set $r_0=b$.
With these assumptions, the attoclock
offset angle in the case of the pure Coulomb potential takes the form
\be
\theta_A=\frac12 \theta\simeq
{\w^2\over E_0^2}{Z\over \rho}
=
{\w^2\over E_0}{Z\over I_p} \ .
\label{KR}
\ee
This equation allows for an alternative interpretation. Indeed, the
Rutherford formula \eref{Rutherford} can be re-written as
\be
\tan{\theta\over2} = {1\over v_\infty\rho}{Z\over v_\infty}=
{1\over L}{Z\over v_\infty} =
{Z\over L}{\w\over E_0}
\ ,
\label{momentum}
\ee
where $L$ is the angular momentum of the projectile. In the attoclock
experiment with circular polarized light, each photon absorption adds
one unit of angular momenta and hence $L\simeq I_p/\omega$. This
immediately leads to  \Eref{KR}.

The signature of this Keldysh-Rutherford (KR) formula is the field
intensity dependence of the offset angle $\theta_A\propto
E_0^{-1}\propto I^{-1/2}$ . This dependence can be understood as a
result of competition of the two terms in \Eref{Rutherford}: the
kinetic energy term $v_\infty^2/2$ and the potential energy term
$Z/\rho$.  As the field intensity grows, the kinetic energy grows
linearly with $I$. Hence this term alone would result in a $I^{-1}$
dependence of the offset angle. This is partially compensated by the
potential energy term as the width of the barrier decreases as
$I^{-1/2}$. Hence the resulting offset angle also decreases as
$I^{-1/2}$.  The signature of the OBI regime is a $I^{-1}$ dependence as
the tunnel width no longer depends on the field intensity.

The recent attoclock measurement on the hydrogen atom
\cite{2017arXiv170705445S} confirmed the $I^{-1/2}$ dependence
experimentally. This measurement, however, used relatively long pulses
without carrier envelope phase stabilization. In addition, the offset angle
extraction procedure was rather complicated due to experimental
constraints. As such this data set does not provide a good reference to test the KR model against. 
Instead, we conduct our own ``numerical experiment'' in a similar manner to that of 
\cite{Torlina2015}. We solve numerically the TDSE 
\begin{equation}
\label{TDSE}
i {\partial \Psi({\bbox r}) / \partial t}=
\left[\hat H_{\rm atom} + \hat H_{\rm int}(t)\right]
\Psi({\bbox r}) \ ,
\end{equation}
where $\hat H_{\rm atom}$ describes the atomic target in the absence of the applied field  
and the interaction
Hamiltonian is written in the velocity gauge
\be
\label{gauge}
\hat H_{\rm int}(t) =
 {\bm A}(t)\cdot \hat{\bm p} \ \ , \ \ 
\bm{E}(t)=-\partial \bm{A}/\partial t \ .
\ee
The vector potential of the driving pulse we take as 
\ba
{\bbox A}(t) &=& 
{-A_0 f(t)}[
\cos(\w t)\hat{{e}}_x
+
\sin(\w t) \hat{{e}}_y
]
 %% \\
 %% {\bbox A}(t) &=& {E_0 f(t)\over \w \sqrt2}[\sin(\w t)\hat
 %%   e_x-\cos(\w t) \hat e_y]
 %% \nn
\label{pulse}
\ea
with the  envelope function
$
f(t) =  \cos^4 (\w t/4) 
$
for $-2\pi/\w<t<2\pi/\w$ and zero elsewhere. 
The (peak) field intensity is given by $I=2(\omega A_0)^2$ and the frequency 
$\omega\simeq 0.057$ a.u.\ corresponds to $800$ nm radiation.
At the tunnelling instant $t=0$ the electric field $E_0$ reaches its
maximum in the $\hat e_y$ direction whereas the vector
potential $A_0$ is largest in the  $-\hat e_x$ direction. The
rotating electric field of the driving pulse causes the photoelectron
to make a single turn by $90^\circ$ before it arrives to the detector
in the $\hat e_x$ direction with momentum $A_0$. Photoelectron scattering in the Coulomb
field adds an offset angle $\theta_A$ relative to this direction (left
panel of \Fref{Fig1}).

\begin{figure}[h]
\hs{-0.5cm}
%\begin{minipage}{0.1\textwidth}
%\hs{-0.6cm}
%\epsfxsize=3.3cm
%\epsffile{Fig2b.eps}
%\vspace{2.5cm}
%\end{minipage}
%\begin{minipage}{0.3\textwidth}
%\hs{0.8cm}
\epsfxsize=5.5cm
\epsffile{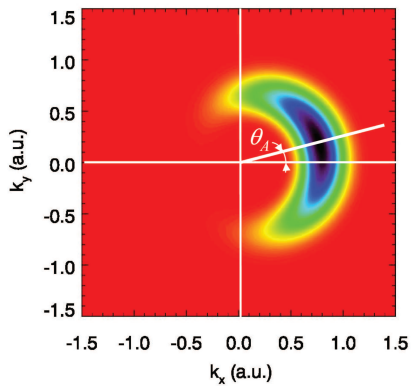}
%\end{minipage}
%\hs{-3mm}
%\def\big{\includegraphics[height=4cm]{Fig2a.eps}}
%\def\little{\includegraphics[height=1.2cm]{Fig2b.eps}}
%\def\stackalignment{l}
%\bottominset{\little}{\big}{5pt}{25pt}
\vs{-3mm}
\caption{Photoelectron momentum distribution in the polarization
  plane at a driving field intensity of $I=$\Wcm{8.6}{13}
  on hydrogen. The offset
  angle $\theta$ relative to the vector potential direction at the
  instant of tunnel is marked. The coloration ranges from zero (red)
  to the maximum amplitude (black) linearly.
\label{Fig2}}

\end{figure}

Solution of the TDSE \eref{TDSE} is found using the iSURF method
implemented in \cite{0953-4075-49-24-245001}.  A typical calculation
takes around 140 CPU hours on a high-performance, distributed-memory cluster. 
The solution of the TDSE is projected on the scattering
states of the target atom thus forming the photoelectron momentum
distribution. A typical 2D momentum distribution in the polarization plane
$k^2P(k_x,k_y)$ is shown on the top panel of \Fref{Fig2}. 
This distribution is integrated radially to obtain the angular distribution
$
P(\theta) = \int dk \ k^2P(k_x,k_y).
$
%
%which is exhibited on the bottom panel of the figure. 
It is then fitted with a Gaussian to determine the peak position and this value
assigned to the attoclock offset angle $\theta_A$. The symmetry of
$P(\theta)$ relative to $\theta_A$ is carefully monitored and serves
as a test of the quality of the TDSE calculation.

\begin{figure}[h]
\hs{-0.5cm}
\epsfxsize=7.5cm
\epsffile{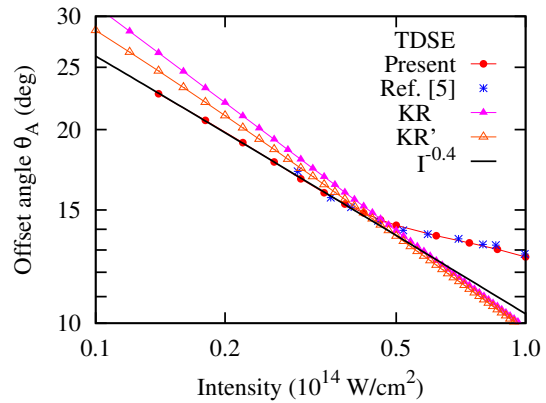}
\vs{-5mm}
\caption{
The attoclock offset angle $\theta_A$ as a function of the field
  intensity $I$ from the present set of TDSE calculations on hydrogen (red filled circles),
the set labeled H2 of \cite{Torlina2015} (blue asterisks) and the KR and KR$'$
models (filled and empty triangles). The present TDSE results are
fitted with $I^{-0.41}$.
\label{Fig3}}
\vspace{-0.4cm}
\end{figure}

Resulting values of the offset angle $\theta_A$ for hydrogen at various field
intensities are plotted in \Fref{Fig3}. We find the present set of TDSE
calculations to be hardly distinguishable from the set labeled H2 of
TDSE calculations reported in \cite{Torlina2015}. This is contrasted with the two
KR and KR$'$ estimates. For the former we simply plot \eqref{KR}, whereas in the latter
we do not make the small angle approximation for the tangent function. 
Accordingly both estimates converge together with
increasing field intensity. The KR scales at all intensities as $I^{-0.5}$ by
construction. Fitting the KR$'$ in the low intesity range yields $I^{-0.44}$.
The TDSE results displays a similar dependency of $I^{-0.41}$
scaling for the same region but then flattens and deviates from
both the KR and KR$'$. This is understandable as the KR model is
expected to be valid only for weak fields where the field driven trajectory
is close to that involved in field-free scattering.

\begin{figure}[h]
\hs{-0.5cm}
\epsfxsize=6.5cm

\epsffile{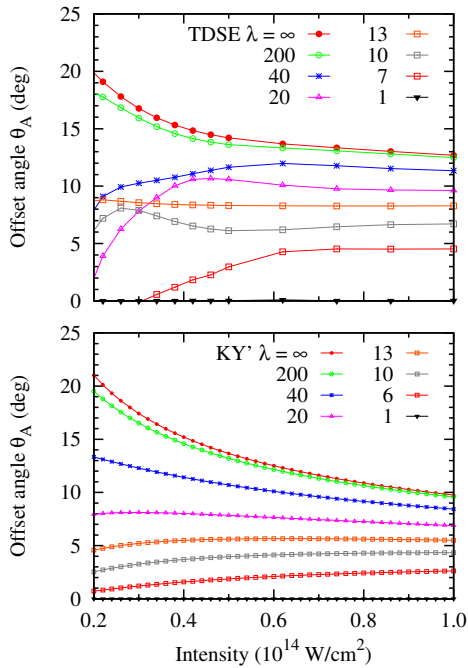}
\vs{-5mm}
\caption{The attoclock offset angle as a function of the field
  intensity for a model Yukawa atom with different screening
  constants $\lambda$. Top: TDSE calculations, bottom: predictions of the
  KY$'$ model.
\label{Fig4}}
\end{figure}

In \Fref{Fig4} we present the offset angles for a model Yukawa atom as a
function of the driving pulse intensity. The top panel contains our TDSE
results whereas the bottom panel displays the predictions of the KY$'$
model, identical to that of KR$'$ but instead based on \Eref{Yukawa}. 
Each curve on these panels correspond to a given value of
the screening constant $\lambda$ which varies from a purely Coulombic case
$\lambda=\infty$, to the severe screening $\lambda=1$ considered in
\cite{Torlina2015,2017arXiv170705445S}. In all cases, the nuclear
charge $Z$ is chosen such that the ionization potential remains
equal to that of the hydrogen atom $I_p=0.5$.  At smaller $\lambda$, the
TDSE results shown on the top panel of \Fref{Fig4} do not follow
predictions of the KY$'$ model faithfully. 
%This is so because the
%scattering in the strongly screened Yukawa potential at lower velocity
%(smaller field intensity in the KY model) is not classical any longer
%\cite{PhysRev.99.1287}.  
This occurs as with decreasing intensity the photoelectron momentum,
and associated de Broglie wavelength, becomes comparable to the
screening length and hence the scattering process can not be
considered classical \cite{PhysRev.99.1287}.  Nevertheless, both sets
of curves are qualitatively very similar.  We observe that the field
intensity dependence of the offset angle for a Yukawa atom is
drastically different from the purely Coulombic case. Instead of the
growth with decreasing field strength, the offset angle becomes
smaller. This is understandable as the tunnel width $b=I_p/E_0$ grows
and becomes comparable with the screening length $\lambda$. In this regime,
the offset angle is exponentially cut off. For the field
intensity of $I=$\Wcm{5}{13}, $b\simeq20$. Accordingly, we observe a
significant decrease of the offset angle at and below this field
intensity for the Yukawa atom with screening length $\lambda=20$. Finally,
the offset angles are essentially zero in both the TDSE and KY$'$ in
the case of severe screening $\lambda=1$ as in
\cite{Torlina2015,2017arXiv170705445S}.

\vs{3.9cm}

\begin{figure}
\epsfxsize=7cm  
\epsffile{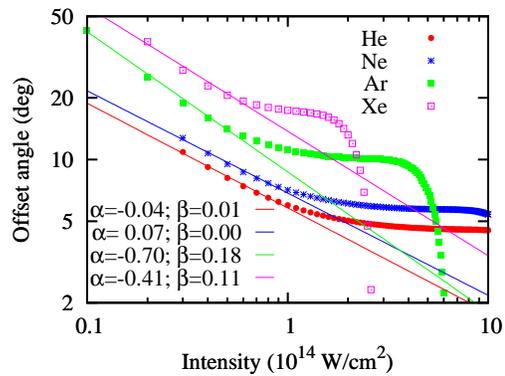}
\vs{-5mm}
\caption{ The attoclock offset angles from CTMC calculations
  \cite{0953-4075-50-5-055602} as functions of the field intensity
  (data points) are fitted with the KR ansatz \eref{fit} (similarly
  colored solid lines).}
\label{Fig5}
\vs{-5mm}
\end{figure}

Although the KR model is developed explicitly for hydrogenic
targets, it can be easily applied to other atoms. Indeed, the
asymptotic charge affecting the departing photoelectron is always the
same for all neutral atomic systems. Hence the basic premise of the KR
model remains valid. To test this model for other atoms, we choose an
extended set of attoclock simulations \cite{0953-4075-50-5-055602}
conducted using the classical-trajectory Monte Carlo (CTMC) method.
We fit the CTMC offset angles with 
 the KR ansatz 
\be
\theta_A(I) =  \frac{\w^2}{ I_p} \frac{(1+\alpha)}{(I/2I_0)^{0.5+\beta}}
 \ ,
\label{fit}
\ee
where $I_0=$\Wcm{3.51}{16} is one atomic unit of field intensity.  We
use $\alpha$ and $\beta$ as fitting parameters which indicate the
deviation of the CTMC calculation from the KR predictions. Results of
the ansatz \eref{fit} application are illustrated in \Fref{Fig5}. We
see that for small intensities the scaling of the offset angles with
the field intensity is indeed close to $I^{-0.5}$. There is most
deviation from the KR prediction in Ar and Xe where the fitting
parameters are comparatively large.  These parameters are in contrast
near zero for the  targets with larger ionization potentials, He and Ne.

In conclusion, we derived a simple empirical expression for the
attoclock offset angle based on the classical Rutherford scattering
formula and the Keldysh estimate of the tunnel width. Utility of these
formulas is demonstrated through application to hydrogen and Yukawa atoms in
addition to noble gas atoms driven by short circularly
polarized pulses.  Due to its simplicity, the Keldysh-Rutherford
formula can be easily applied to attoclock experiments with arbitrary
polarization though modification of the above formalism to account 
for non-unitary ellipticity. 
One such case being the recent attoclock measurements on
atomic hydrogen \cite{2017arXiv170705445S}, where the signature field
intensity scaling of the KR model $I^{-0.5}$ was indeed observed.
These measurements as well as the earlier ``numerical experiment''
\cite{Torlina2015} were indicative of a zero
tunnelling time. This interpretation followed from the TDSE
simulations with the screened Yukawa potential with $\lambda=1$ which set
the attoclock offset angles to zero independently of the field
intensity. In the present work we show that this interpretation is
overly simplistic. Indeed, the cross-over from the hydrogen to Yukawa atoms
is smooth and by decreasing the screening length $\lambda$ one can switch
off the action of the Coulomb field gradually. In fact it is only the shortest
screening length $\lambda=1$ considered here that sends the offset angles below
$1^\circ$. Nonetheless, in each case we find the offset angles to follow closely
the predictions of the KR model and as such are not, most likely, related
to the time the tunnelled electron spends under the barrier.

The authors are very thankful to Satya Sainadh, Nicolas Duguet, Igor
Ivanov, Klaus Bartschat, Igor Litvinyuk and Robert Sang for many
stimulating discussions.  The authors are also greatly indebted to
Serguei Patchkovskii who placed his iSURF TDSE code at their disposal.
Finally we thank Zengxiu Zhao for supplying the data
\cite{0953-4075-50-5-055602} in numerical form.  S.E. was
supported by the DFG Priority Programme DO 604/29-1 and
A.K.\  by the Wilhelm and Else Heraeus Foundation.
Resources of the National Computational Infrastructure were employed.

\np~~\np

%\bibliography{references,areferences,ereferences,sreferences,ireferences,treferences,mypapers}

\end{document}